**Coauthorship networks: A directed network approach considering the order and number of coauthors**

**Authors**: Jinseok Kim and Jana Diesner

**Author Information:**

Jinseok Kim
Graduate School of Library and Information Science, University of Illinois at Urbana-Champaign

Jana Diesner
Graduate School of Library and Information Science, University of Illinois at Urbana-Champaign

**Corresponding Author:**
Jinseok Kim, jkim362@illinois.edu

**Abstract**

In many scientific fields, the order of coauthors on a paper conveys information about each individual's contribution to a piece of joint work. We argue that in prior network analyses of coauthorship networks, the information on ordering has been insufficiently considered because ties between authors are typically symmetrized. This is basically the same as assuming that each co-author has contributed equally to a paper. We introduce a solution to this problem by adopting a coauthorship credit allocation model proposed by Kim and Diesner (2014), which in its core conceptualizes co-authoring as a directed, weighted, and self-looped network. We test and validate our application of the adopted framework based on a sample data of 861 authors who have published in the journal Psychometrika. Results suggest that this novel sociometric approach can complement traditional measures based on undirected networks and expand insights into coauthoring patterns such as the hierarchy of collaboration among scholars. As another form of validation, we also show how our approach accurately detects prominent scholars in the Psychometric Society affiliated with the journal.

**Keywords**: social network analysis, coauthorship network, coauthor order, bibliometrics, sociology of science

**Introduction**

The increase of coauthored research publications has led to a need for a better understanding of the fundamental principles of scientific collaboration (He, Ding, & Yan, 2012; Wray, 2002). The order of coauthors is one aspect of research on this broader topic that has particularly significant practical implications: authorship information is used to assess scholars for employment, funding and tenure. The underlying assumption here is that the order of coauthors implies information about the contribution of the involved individuals to a project, and the credit they deserve for it (Beasley & Wright, 2003; Thomas et al., 2004). There is a large body of work on investigating and modeling the conventions for ordering coauthors across domains and on inferring the amount of each author's contribution from the coauthor order (for a comprehensive review see Marušić, Bošnjak, & Jerončić, 2011).

While coauthorship has also been heavily studied by network scholars (Barabasi et al., 2002; Goyal, van der Leij, & Moraga-Gonzalez, 2006; Moody, 2004; Newman, 2001), the ordering of coauthors has hardly been addressed from a network analytic perspective. In most coauthorship network studies, coauthoring relationships are conceptualized as undirected and binary (sometimes weighted) graphs (De Stefano, Giordano, & Vitale, 2011). Since ordering effects are not considered in previous research, their implied assumption would be that coauthors contribute equally to a paper.

This procedure may conflict with the argument made by various scholars that, in order to measure the impact of authors in an objective fashion, their ordering needs to be considered (Jennings & El-adaway, 2012; Wren et al., 2007). In line with this thinking, we propose a method that accounts for author ordering with the ultimate goal of contributing to a more holistic understanding of the structure and implication of scholarly collaboration. In the following sections, we review previous coauthorship network studies in terms of coauthor order. Then, we introduce a framework conceptualizing the coauthorship network as a directed, weighted, and self-looped sociometric graph, which has been proposed by Kim and Diesner (2014). We provide an empirical example to illustrate and evaluate the application of the proposed method. Finally, we discuss outcomes with limitations and future directions.

**Background**

Coauthorship networks have been intensively studied in many fields; mainly with a focus on macro-level properties of the networks. For example, multiple studies have confirmed the power-law distribution of the number of collaborators per author and the small-world structure of coauthor networks (Barabasi et al., 2002; Liu, Bollen, Nelson, & Van de Sompel, 2005; Newman, 2004; Rodriguez & Pepe, 2008). Other studies identify actor-level characteristics such as the centrality of individual scholars (Ding, 2011; Yan & Ding, 2009). Both research traditions, however, have not paid much attention to the order of coauthors. One of the reasons for this might stem from a methodological convention (Faust, 1997): in coauthor networks, two authors are connected if they have worked together on a publication. Such a network can be represented as an adjacency matrix, where authors are denoted in the rows and publications in the columns. These two-mode or bipartite graphs are then usually folded into an author-by-author matrix for coauthorship network analysis. This method of network construction inevitably symmetrizes all relationships (see Figure 1), where only the presence or absence of a relation matters (Barabasi et al., 2002; Newman, 2001; Moody, 2004). Several scholars have attempted to advance this traditional approach. For example, some have assigned weights to undirected ties according to a) the inverse of the number of authors per publication and b) the cumulative frequency of collaboration between any pair of authors (De Stefano et al., 2011; Fiala, Rousselot, & Jezek, 2008; Newman, 2004; Sidiropoulos & Manolopoulos, 2006; Yan & Ding, 2011).

Figure 1: Undirected, Binary or Weighted Approach

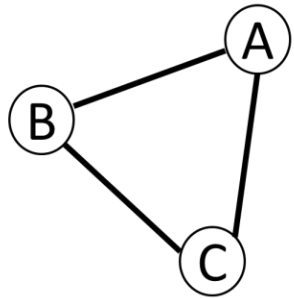

Others have turned to directed networks for developing sophisticated measures of author centrality. For example, Liu and colleagues (2005) modeled a directed coauthorship network by replacing all undirected relations with directed, reciprocated relations (see Figure 2). Yoshikane and colleagues (2006, 2009) represented the order of coauthors as a directed network, where only the first author receives ties from coauthors (see Figure 3).

Figure 2: Directed, Reciprocated Network Approach

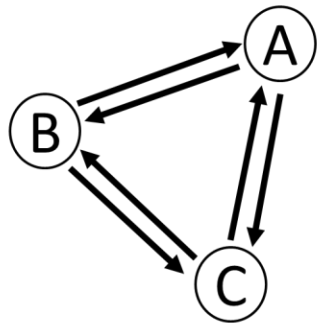

Figure 3: Directed, 1st Author-Only Network Approach

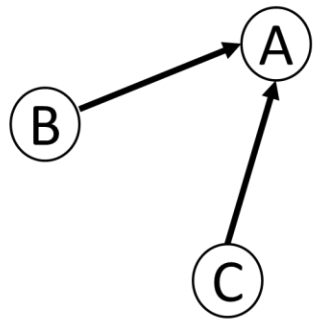

Despite these advances, the common problem with most prior studies is that, given the symmetric nature or lack of hierarchies in coauthor relationships, ordering is not represented and considered. In the case of the directed network conceptualization by Liu and colleagues (2005), every pair of coauthors is connected via reciprocal ties. Even in Yoshikane et al.'s studies (2006, 2009), where the coauthor order is considered, coauthors except the first author are disconnected from each other, and their order is ignored.

The approaches discussed above - except the one by Yoshikane et al. (2006, 2009) - are appropriate when being applied to coauthorship networks where most of the coauthors are ordered alphabetically, which reflects equal contributions of all involved scholars, as often in economics and mathematics (Endersby, 1996; Laband & Tollison, 2006; Riesenberg & Lundberg, 1990). The majority of scientific fields, however, have been reported to represent the coauthor order according to authors' relative contribution (He et al., 2012; Spiegel & Keith-Spiegel, 1970; Wagner, Dodds, & Bundy, 1994; Waltman, 2012). Therefore, the aforementioned coauthorship network approaches, when applied to these fields, might cause a loss of information and lead to biased or false findings. We address these issues by adopting a method proposed by Kim and Diesner (2014) for modeling coauthorship networks that explicitly considers the order of coauthors. In the next section, we introduce this framework and explain how it overcomes the limitations of prior work on this topic.

**Methodology**

*Conceptualization of a Directed Coauthorship Network*

Synthesizing prior studies, we argue that the following three desiderata ought to be met in order to appropriately represent coauthor networks.

D1: All authors of a paper should be connected. This connectedness conveys the idea that authors are involved with one another in collaboration. Most previous coauthorship studies have assumed the undirected connectedness among collaborators.

D2: Coauthor order should be explicitly represented in the network data. This desideratum is directly related to the hierarchical arrangement of coauthors: the first author should come before the second, the second before the third, and so on.

D3: Each author's individual contributions should be scaled by both her rank in the coauthor list and the number of authors on a publication. Although in some fields coauthors are acknowledged with the same maximum credit for a single piece of work, i.e. one publication per each coauthor (Chan, Chen, & Cheng, 2009), it is common that authors are given less credits when the number of coauthors becomes larger (Newman, 2004; Wren et al., 2007).

In order to combine these three requirements, we turn to Kim and Diesner (2014)'s work, where coauthorship credit allocation is conceptualized by a directed network approach. Their Network-Based Allocation (NBA) model of coauthorship credit is based on the following assumptions:

A1: Coauthors determine the order of coauthors according to their relative contribution to a paper. The amount of contribution decreases from the first to the last author.

A2: Each author is given an initial coauthorship credit that represents the unit value per publication divided by the number of coauthors. Here, each paper is assumed to have an equal value of one. This parameter setting is based on previous studies (Galam, 2011; Hagen, 2010; Vinkler, 1993).

A3: Coauthors of a paper distribute a certain proportion of their initial coauthorship credit in equal amounts to their preceding coauthors.

Building upon these assumptions, NBA models the situation of scholars being coauthors as an author transferring a part of her coauthorship credit to other authors as a sign of acknowledgement of their contribution. The application of this conceptualization to the three-coauthor case is shown in the sociometric digraph below (see Figure 4).

Figure 4: Three Coauthors Relationship Visualized by a Directed Network

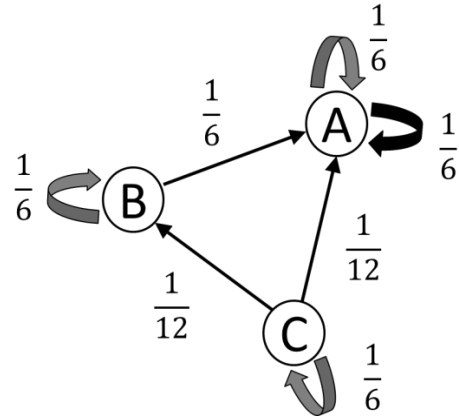

In the shown diagraph, the respective order of authors is reflected by the number of incoming and outgoing ties (except for self-loops). Lead author A receives two ties from both authors B and C. Author B receives one tie from C while sending one to A. Author C sends two ties without receiving any tie. This inequality of exchange of directed ties generates a hierarchy between authors: author A precedes author B who is followed by author C in the coauthor list of a three-authored paper.

Each coauthor is given an initial coauthorship credit (IC), the paper value (assumed to be 1) divided by the size of collaborators (3 in Figure 4). Thus, IC here is 1/3. After coauthors are given IC, they distribute a part of their IC (= transferable credit, TC) in equal portions to others preceding them in terms of the coauthor order while keeping the rest (= non-transferable credit, NC) to themselves. Thus, the initial coauthorship credit equals the sum of the transferable and the non-transferable credits (i.e., IC = TC + NC).

The transferable credit (TC) is calculated by multiplying a distribution factor ($d$) to IC. Here, $d$ can be any real number between zero and one, representing the ratio of IC that should be disseminated by each coauthor. In the example above, half of the authors' ICs ($d = 1/2 = 0.5$) are distributed such that 1/6 (i.e., TC = 1/2 × 1/3) of each authors' credit is transferred. Here, the last author C has to divide her TC into two portions (TC × 1/2 = 1/12) in order to equally distribute to author A and B, while she keeps the rest (i.e., NC = IC − TC = 1/3 – 1/6 = 1/6). The second author B transfers half of her IC (i.e., TC = 1/6) to the first author A and keeps the rest. Author A has no preceding author such that all of her TC is allocated to herself (the black, self-looped tie in Figure 4). She also keeps the NC. In the end, author A holds a total of 7/12 credits for the paper, author B holds 3/12 and author C has 1/6. Each author keeps the non-transferable credit (NC =1/6) to themselves, which is depicted by a gray-colored, self-looped tie originating from each author and directed towards each node themselves in Figure 4. This relational allocation of authorship credit among coauthors can be generalized and applied to any multi-authored paper.

Noticeably, the adopted model produces different authorship credits even for the same paper depending on the distribution factor ($d$), which equips the model with flexibility in allocating coauthorship credit scores. For example, the first author in a three-authored paper can receive a minimum credit of 0.33 ($d = 0$) and a maximum of 0.83 ($d = 1$) depending on $d$. Moreover, distribution factors can be assigned different values according to the number of coauthors. For instance, a two-authored paper can be modeled with a $d$ different from that assigned to a three-authored paper. We refer readers to Kim and Diesner (2014) for more detailed explanation on the NBA model.

So far, we have illustrated that ordered coauthor relationships can be conceptualized into a directed, weighted, and self-looped network according to NBA model proposed by Kim and Diesner (2014). Such a conceptualization, i.e., a directed, weighted coauthorship network, however, is not new. West, Jensen, Dandrea, Gordon, and Bergstrom (2013) modeled authors' relationship through a directed, weighted network in citation analysis. Li and You (2013) conceptualized a directed, weighted, and self-looped coauthorship network where an author's "energy" flows. But those models are different from the NBA model in that they still assume an equal contribution among coauthors. The uniqueness of NBA model lies in that this conceptualization integrates the connectedness of coauthors, their order, the number of collaborators, and each author's contribution into one framework, which together satisfy the desiderata proposed above. Thus, we adopt this NBA model and extend its application to coauthorship network analysis.

*Measures for coauthorship network*

The aforementioned conceptualization of coauthorship networks enables us to compute new metrics that supplement prior metrics used in previous coauthorship network studies. One primary use of network analysis in coauthorship research is the identification of prominent actors in the network. According to Knoke and Burt (1983) and Wasserman and Faust (1994), prominence is defined as a greater visibility of an actor compared to others. They define two general types of prominence: centrality for symmetric (undirected) relations and prestige for asymmetric (directed) ones. Drawing on this categorization and the previous works applying centrality measures to coauthorship analysis (Liu et al., 2005; Yan & Ding, 2009; (Yin, Kretschmer, Hanneman, & Liu, 2006)), we apply four prominence measurements to the coauthorship network analyzed herein: degree, betweenness, and closeness centralities and indegree prestige, as defined in Table 1.

Table 1: Overview of Selected Prominence Measures ($g$ is the total number of actors in a network)

**Degree Centrality**

$$C_D(i) = \sum_{j=1}^{g} x_{ij} (i \neq j)$$

$x_{ij}$ represents the presence of relationship between actor $i$ and $j$

**Betweenness Centrality**

$$C_B(i) = \sum_{j,k \neq i} \frac{n_{jk}(i)}{n_{jk}}$$

$n_{jk}$ is the number of shortest paths (geodesics) between actor $j$ and $k$, and $n_{jk}(i)$ is the number of geodesics between actor $j$ and $k$ that includes actor $i$

**Closeness Centrality**

$$C_c(i) = \sum_{j=1}^{g} \frac{1}{d(i,j)}$$

$d(i,j)$ is the distances between actor $i$ and $j$

**Indegree Prestige**

$$P_I(i) = \sum_{j=1}^{g} x_{ji}$$

$x_{ji}$ is the number/value of a tie directed from actor $j$ to actor $i$

Centrality is concerned with an actor's relationships with others (Freeman, 1978; Wasserman & Faust, 1994). Specifically, in a coauthorship network, the relation between a pair of actors represents the fact that these individuals are associated with each other through collaboration (Shumate et al., 2013). Directionality does not apply in this case. Thus, centrality measures have been defined and used mostly in coauthorship studies where undirected relations are assumed.

In an undirected, binary coauthorship network, degree centrality measures the number of unique coauthors that an author has with no sensitivity to the number of joint publications (Barabasi et al., 2002; Moody, 2004; Newman, 2001). Here, an author is central when she has many collaborators. In an undirected coauthorship network, an author with high betweenness centrality has an advantage over others in terms of connecting diverse authors from different affiliations or domains, or controlling the flow of information on collaboration (Newman, 2001; Yan & Ding, 2009). As a distance-based measure like betweenness centrality, closeness centrality in a coauthoring relationship measures the extent to which an author can be easily connected to all the other authors in a network, and thus can mobilize the network without depending much on intermediary coauthors (Prell, 2012). One shortcoming of closeness centrality (Freeman, 1979) in this context is that it cannot be computed on a disconnected network, and most of coauthorship networks are disconnected. To avoid this problem, this study calculates closeness centrality by summing the reciprocal distances between all actors (Borgatti, 2006).

Unlike centrality measures, prestige is mainly concerned with the direction of a relationship. Prestige refers to the extent to which an actor becomes a recipient or target of relations initialized by others in the network (Knoke & Burt,

1983). Specifically, an actor's prestige in a network increases when she receives many nominations from other actors. Based on this line of argumentation, prestige cannot be measured in an undirected network (Wasserman & Faust, 1994). As shown in the conceptualization section above, we model coauthorship networks as directed graphs, where an author transfers a portion of her coauthorship credit to her coauthors as an acknowledgement or endorsement of their contribution. This conceptualization necessitates a directed flow network instead of an undirected, representational coauthorship network (Shumate et al., 2013). Therefore, the coauthorship networks as modeled in this study can leverage existing prestige metrics, which has been inapplicable with previous undirected network approach.

An author in a directed coauthorship network is expected to receive more credits from coauthors when she is often fairly upfront in the coauthor order and/or when the number of collaborators is small. Thus, a prestigious author in terms of indegree prestige can be said to have led collaboration more often than other scholars and/or with a small number of collaborators per paper.

In the next section, a coauthorship network of scholars in a psychological journal will be processed for illustrating the application of conceptualization of directed coauthorship network. The selected set of prominence measures is used to produce rankings of authors. We compare the resulting rankings against each other as well as to real world data on prominent scholars in the field related to the journal.

**Analysis**

*Data Processing*

In this section, we report on the testing of the proposed solution in a real-world application setting. We compare our results to a proxy of ground truth – i.e. expert verified data. For this purpose, we use a sample dataset of coauthors who published in Psychometrika, one of leading journals in quantitative psychology (Burgard, 2001). We obtained the data from two databases: PsycINFO and Scopus. PsycINFO has consistently maintained full names of authors, which are helpful to disambiguate author names. Scopus logs the records of corresponding/reprint author. For querying these databases, we used "Psychometrika" as "Publication Name" and further constrained the search to "Article" as "Document Type". We retrieved 1,263 (PsycINFO) and 1,208 (Scopus) articles published between 1980 and 2012. Both datasets were combined and de-duplicated, resulting in 1,161 unique articles. Among these articles, 498 articles (43%) had one author, while 663 articles (57%) are multi-authored. As we are interested in coauthoring relation, we analyzed the 663 multi-authored articles. Several authors use middle names or initials in some articles, but not in others. We manually disambiguated and consolidated names by leveraging affiliation or correspondence information provided in the articles as well as public information on these people. This reference resolution step identified 861 unique authors.

Our approach allows for unequal contributions of coauthors. In the field of psychology, authorship order reflects the relative magnitude of individual contribution (Maciejovsky, Budescu, & Ariely, 2008; Spiegel & Keith-Spiegel, 1970), with the first author typically having contributed the most (Maciejovsky et al., 2008). In some fields the last author is regarded as the senior author in charge of the paper. Scholars have reported different conventions on the last author's contribution to joint work. Sometimes, the last author is viewed as contributing as much as the first author (Jian & Xiaoli, 2013). Other times, she is regarded as contributing less than the first author but more than the other coauthors (Retzer & Jurasinski, 2009; Tscharntke, Hochberg, Rand, Resh, & Krauss, 2007). The contribution of the last author in the field of psychology is not reported on in prior scientometric research.

To address this limitation, we use the information about the corresponding author as a proxy of indication of contribution (Milojević, 2012; Wren et al., 2007). Most journals require one of the coauthors to be identified as the corresponding author (or reprint author), who is usually regarded as a principal investigator, project leader, or mentor of graduate students (Jian & Xiaoli, 2013). The corresponding author is often considered as the most important contributor to a collaborative project (Mattsson et al., 2011). In Psychometrika, 87.3% of 663 multi-authored papers list the first author as the reprint or corresponding author, while 10.2% specify the last author and 2.5% the middle authors. This empirical finding validates our conceptualization of the first author as the lead contributor. This decision is in sync with prior research in psychological scientometrics, which confirmed the same effect (Maciejovsky et al., 2008). Based on this rationale, we assume that the last author contributes the least when the first and corresponding authors are the same. In cases where the first and corresponding authors are different, we assume the corresponding author to be the lead contributor (Mattsson et al., 2011; Wren et al., 2007)[1]. We practically implement this choice by rearranging the list of authors: the corresponding author gets placed first, followed by the original first author and consecutively by all other coauthors.

*Fitting Model to Empirical Data*

Another issue that arises when applying the new model to coauthorship data is how to decide on the amount of coauthorship credit assigned to each author. Maciejovsky and his colleagues (2008) analyzed the order of coauthors and the perceived contribution per author in a study with 52 professors and graduate students in psychology. Researchers presented the respondents with a total of 1,702 lists of coauthors of two, three, and four-coauthored ($N$) papers, and asked them to assign the amount of contribution (from 0% to 100%) to each coauthor according to their rank ($r$) in the coauthor order (see Table 2). The outcome of this study was used as a proxy of ground-truth data to estimate the best distribution of contribution size of coauthors in Psychometrika[2].

Table 2: Coauthorship Credit Share of *r*-th Author in an *N*-authored Paper in Psychology (Maciejovsky et al., 2008)

| $N$ | 2 | | 3 | | | 4 | | | |
|---|---|---|---|---|---|---|---|---|---|
| $r$ | 1 | 2 | 1 | 2 | 3 | 1 | 2 | 3 | 4 |
| Credit Share | 0.61 | 0.39 | 0.49 | 0.29 | 0.22 | 0.42 | 0.24 | 0.19 | 0.14 |

In our model, the distribution factor ($d$) affects the amount of contribution per author. To incorporate the coauthorship credit allocation information from Maciejovsky et al. (2008) into our study, we identified the distribution factor that best fits the empirical data. This was done by calculating the Lack of Fit (LOF) as a standardized deviation from model predictions as follows:

$$LOF = \frac{1}{(n-1)} \sum \frac{(E-C)^2}{C} \qquad (1)$$

Here, $n$ denotes the total number of empirical observations, $E$ the empirical data based on the proxy of ground-truth data and $C$ the scores generated by our model. For example, to find the best fitting set of contribution scores from the model for the two coauthored case, the proxy of ground-truth values of 0.61 and 0.39 from the Maciejovsky et al. (2008) study were selected. Then, those two values were compared to the scores obtained with our model run with various distribution factors ranging from 0.0 to 1.0. From the obtained results we selected the distribution factor with the lowest LOF value. Here, the lowest LOF represents the minimum error or difference between the proxy of ground-truth data and the model-generated scores. Our analyses show that the distribution factor of $d = 0.21$ best fits the proxy of ground-truth scores of two coauthors case, $d = 0.33$ the three coauthor case, and $d = 0.39$ the four coauthor case (see Table 3). This implies that, as the distribution factor increases with the coauthor size, the coauthorship credit allocation is less equal in psychology as more coauthors are involved.

The Psychometrika data includes papers authored by two to twelve people, while the proxy of ground-truth data study only covers papers with two to four coauthors. Notably, papers written between two to four people constitute almost 99% of the dataset. We decided to use the distribution factor for four coauthor case ($d = 0.39$) to the papers with more than four coauthors.

Table 3: The Model's Coauthorship Credit Share Fitted to Empirical Data from Maciejovsky et al. (2008)

| N | 2 | | 3 | | | ≥ 4 | | | |
|---|---|---|---|---|---|---|---|---|---|
| r | 1 | 2 | 1 | 2 | 3 | 1 | 2 | 3 | 4 |
| Empirical Data | 0.61 | 0.39 | 0.49 | 0.29 | 0.22 | 0.42 | 0.24 | 0.19 | 0.14 |
| Model Score | 0.61 | 0.40 | 0.50 | 0.28 | 0.22 | 0.43 | 0.23 | 0.19 | 0.15 |
| d | 0.21 | | 0.33 | | | 0.39 | | | |
| LOF | 0.000001 | | 0.000528 | | | 0.000297 | | | |

In Figure 5, the coauthorship credit shares generated by the model (with black-diamond points) are compared to those from the proxy of ground-truth empirical data (with grey-circle points, hard to see when the black and grey lines are on top of each other) for two to four coauthor cases (denoted by $N$). In that figure, the x-axes represent author ranks, and the y-axes represent coauthorship credit shares. According to the results, the models' scores seem to be overall in close agreement with the proxy of ground-truth data. This also means that the model of our study generates a distribution of coauthorship scores that resemble it.

Figure 5: The Model's Credit Scores Compared with Empirical Data for $N$-authored Papers (x-axes: author ranks, y-axes: credit scores)

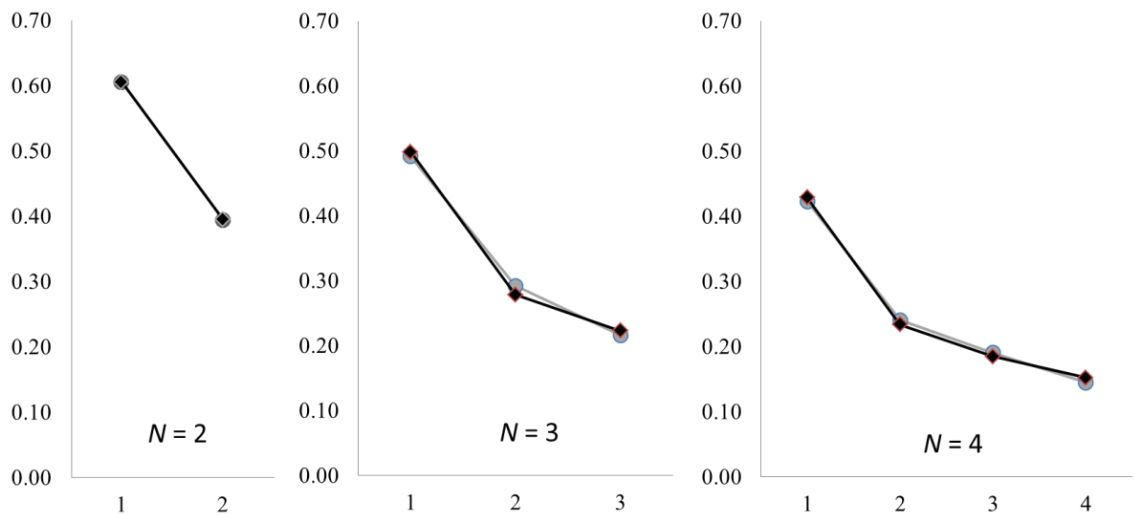

*Undirected vs. Directed Network Construction*

To assess the performance of the proposed approach, we compare the obtained results to those from alternative prior approaches to analyzing coauthorship networks. The coauthorship network of Psychometrika was first generated as a directed, weighted, and self-looped network according to our conceptualization. Here, the weight of directed ties was assigned by the model to produce each author's coauthorship credit share that best fits the proxy of ground-truth data from Maciejovsky et al. (2008). This network was used for calculating indegree prestige. Then, the network was symmetrized as done in most previous studies. With this undirected network approach, the coauthor order is ignored. This undirected network was used for degree, betweenness, and closeness centrality measures.

**Results**

*Descriptive Statistics*

Our study considered 861 unique authors in the Psychometrika coauthorship network. A paper has on average 2.41 authors, and the average author collaborates with 2.52 other scholars. The graph is not connected: the largest component has 360 authors (42% of all coauthors), the second largest one has 14 scholars. The overall density of the network is 0.003.

Table 4: Summary Statistics for Coauthoring in Psychometrika

| | |
|---|---|
| Total of multi-authored papers | 663 |
| Two-coauthored | 464 |
| Three-coauthored | 144 |
| Four-coauthored | 47 |
| Five-coauthored | 4 |
| Six-coauthored | 2 |
| Seven-coauthored | 1 |
| Twelve-coauthored | 1 |
| Number of unique authors | 861 |
| Average number of papers per author | 1.86 ($SD = 2.65$) |
| Average number of authors per paper | 2.41 ($SD = 0.79$) |
| Average number of coauthors per author | 2.52 ($SD = 2.70$) |

*Rankings of Authors*

Individual authors can be ranked according to different dimensions of prominence (see Table 5). Each measure results in a different list, and each metric sheds light on a different aspect of scientific collaborations in these data.

Table 5: Top 20 Authors per Prominence Measures

| | Centrality | | | Prestige |
|---|---|---|---|---|
| Rank | Degree | Closeness | Betweenness | Indegree |
| 1 | desarbo, wayne s. | desarbo, wayne s. | verhelst, norman d. | ten berge, jos m. |
| 2 | de boeck, paul | takane, yoshio | bentler, peter m. | bentler, peter m. |
| 3 | ten berge, jos m. | de leeuw, jan | heiser, willem j. | lee, sik-yum |
| 4 | takane, yoshio | ten berge, jos m. | de leeuw, jan | kiers, henk a. l. |
| 5 | bentler, peter m. | kiers, henk a. l. | takane, yoshio | desarbo, wayne s. |
| 6 | boker, steven m. | carroll, j. douglas | desarbo, wayne s. | yuan, ke-hai |
| 7 | carroll, j. douglas | bentler, peter m. | groenen, patrick j. f. | takane, yoshio |
| 8 | de leeuw, jan | mooijaart, ab | kiers, henk a. l. | brusco, michael j. |

| 9 | heiser, willem j. | hwang, heungsun | de boeck, paul | van mechelen, iven |
| 10 | kiers, henk a. l. | heiser, willem j. | ten berge, jos m. | de leeuw, jan |
| 11 | chow, sy-miin | de soete, geert | van der heijden, peter g. m. | van der linden, wim j. |
| 12 | hwang, heungsun | groenen, patrick j. f. | sijtsma, klaas | de boeck, paul |
| 13 | lee, sik-yum | verhelst, norman d. | de soete, geert | hwang, heungsun |
| 14 | sijtsma, klaas | kroonenberg, pieter m. | zhang, guangjian | carroll, j. douglas |
| 15 | van der linden, wim j. | young, forrest w. | jennrich, robert i. | ceulemans, eva |
| 16 | van mechelen, iven | van der heijden, peter g. m. | mislevy, robert j. | meredith, william |
| 17 | bates, timothy | de boeck, paul | yuan, ke-hai | heiser, willem j. |
| 18 | brick, timothy | bekker, paul a. | chow, sy-miin | böckenholt, ulf |
| 19 | estabrook, ryne | satorra, albert | mooijaart, ab | ramsay, james o. |
| 20 | fox, john | furnas, george w. | zhang, zhiyong | sijtsma, klaas |

To give a more detailed illustration, Table 6 shows the degree centrality and indegree prestige rankings for two authors who each wrote 11 articles. Although they have the same level of productivity in terms of number of papers published, they differ from a network analytical perspective. For example, according to the number of unique collaborators (degree centrality), Carroll ranks higher than Hwang because he has 13 unique coauthors, while Hwang has 12.

Table 6: An Illustrated Comparison of Two Prominence Rankings of Two Authors

| Measure | Carroll, J. Douglas | Hwang, Heugnsun |
|---|---|---|
| Number of papers | 11 | 11 |
| Degree Ranking | 8 | 13.5 |
| Indegree Ranking | 14 | 13 |

According to the indegree prestige measure, however, Hwang ranks higher than Carroll. The indegree measure favors an author who often leads coauthoring and/or publishes single papers with a small number of coauthors. As shown in Table 7[3], Hwang served as the first author more often than Carroll did. The amount of coauthorship credits that Hwang received - according to the directed coauthorship network model of this study – is 4.96, while Carroll received 4.18.

Table 7: Coauthor Lists of Two Authors (names in bold and coauthor names separated by semicolons)

| Co-author Lists of Papers **J. Douglas Carroll** Participated In |
|---|
| **carroll, j. douglas** ; winsberg, suzanne |
| de soete, geert ; **carroll, j. douglas** |
| **carroll, j. dougla**s ; arabie, phipps |
| takane, yoshio ; **carroll, j. douglas** |
| desarbo, wayne s. ; **carroll, j. douglas** |
| arabie, phipps ; **carroll, j. douglas** |
| **carroll, j. douglas** ; pruzansky, sandra ; kruskal, joseph b. |

| Co-author Lists of Papers **Heungsun Hwang** Participated In |
|---|
| takane, yoshio ; **hwang, heungsun** |
| **hwang, heungsun** ; takane, yoshio |
| **hwang, heungsun** ; takane, yoshio |
| **hwang, heungsun** ; takane, yoshio |
| **hwang, heungsun** ; dillon, william r. ; takane, yoshio |
| **hwang, heungsun** ; desarbo, wayne s. ; takane, yoshio |
| **hwang, heungsun** ; ho, moon-ho ringo ; lee, jonathan |
| takane, yoshio ; **hwang, heungsun** ; abdi, hervé |
| jung, kwanghee ; takane, yoshio ; **hwang, heungsun** ; woodward, todd s. |
| **hwang, heungsun** ; jung, kwanghee ; takane, yoshio ; woodward, todd s. |
| **hwang, heungsun** ; suk, hye won ; lee, jang-han ; moskowitz, d. s. ; lim, jooseop |

weinberg, sharon l. ; **carroll, j. douglas** ; cohen, harvey s.

pruzansky, sandra ; tversky, amos ; **carroll, j. douglas**

desarbo, wayne s. ; **carroll, j. douglas** ; clark, linda a. ; green, paul e.

de soete, geert ; desarbo, wayne s. ; furnas, george w. ; **carroll, j. douglas**

*Rank Order Correlation*

As shown above, each prominence measure ranks authors differently. For the macroscopic comparison of prominence metrics, we used Kendall's tau ($\tau$) rank order correlation. Kendall's tau was chosen over Spearman's rank order correlation because each measure produces many tied ranks, especially for mid and low ranked authors. Kendall's tau is a non-parametric correlation measure which is useful when comparing dataset with many tied ranks (Field, 2009). The rankings of the three prominence measures employed herein were compared against each other (see Table 8). In the comparison matrix, significance is signaled by "*" (at the 0.01 level, 2-tailed), assigned to the right- upper side of each correlation coefficient value.

Table 8. Kendall's Tau Test between Prominence Measures

|  | Degree | Betweenness | Closeness | Indegree |
|---|---|---|---|---|
| Degree | - | .524* | .489* | .008 |
| Betweenness |  | - | .346* | .517* |
| Closeness |  |  | - | .037 |
| Indegree |  |  |  | - |

A noticeable feature here is the lack of dependency between indegree prestige and degree centrality ($\tau$ = .008, non-significant) and between prestige and closeness ($\tau$ = .037, non-significant). This shows that these measures capture different aspect of coauthoring in Psychometrika. Authors working with many collaborators are not necessarily those who collaborate more often or contribute more to collaboration than others. This might be partly due to the fact that degree centrality inflates the importance of an author if she participates in a paper with many coauthors. The betweenness centrality consistently shows an intermediate level of correlation with both degree ($\tau = .489$) and indegree ($\tau = .517$).

*Hierarchy of collaboration*

The directed coauthorship network in this study is based on the hierarchical arrangement of coauthors per paper, from which we infer their relative contribution. To explore the overall hierarchical structure of coauthorship network in Psychometrika, authors were divided into three Blocks according to the indegree prestige ranking: top 20% (Block1: rank 1 to 175), middle 30% (Block2: rank 176 to 432), and lower 50% (Block3: rank 433 to 861) groups. We have no theoretical grounds for grouping scholars in cases like our study. Therefore, we first divided authors in two groups: upper 50% and lower 50%. After that, we additionally divide the upper group into top 20% and the rest 30% because we might expect to see the so-called '20:80' distribution. Then, their credit transfer was aggregated into the block level, and the coauthorship network was collapsed into a block matrix to represent the movement of credit between blocks (see Table 9).

Table 9: Coauthorship Credit Transfer between Blocks

|        | Block1 | Block2 | Block3 |
|--------|--------|--------|--------|
| Block1 | 349.24 | 8.72   | 2.43   |
| Block2 | 8.62   | 132.47 | 0.90   |
| Block3 | 18.38  | 16.35  | 125.88 |

In table 9, blocks listed in the rows send credits to blocks in the columns. For example, the first row tells that Block1 sends credits to Block2 (8.72) and Block3 (2.43). The diagonal represents the transfer among within-block members. For example, in the first row, a total of 349.24 credits are transferred among authors belonging to Block1. This table shows that 92% of credit transfer occurs in the diagonal (grey-shaded) [4]. This is mainly due to the fact that the diagonal includes (1) the non-transferable credits assigned to each author before any transfer happens, which accounts for almost 75% of all credits (= 496.59/663.00), (2) the self-allocated credits of first authors, and (3) the credits transferred among authors. To make the table more interpretable, these three types of credits in the diagonal need to be considered separately as shown below.

First, Table 10 looks at transfer relationship from a different angle: here, the non-transferable credits were removed from the diagonal of the credit mobility matrix. In the coauthorship network model of this paper, each author in a paper (a) receives an initial coauthorship credit (= the unit value of a paper divided by the number of coauthors), (b) distributes a part of the initial credit in equal amounts to other coauthors preceding him in coauthor order (= transferable credit decided by a distribution factor), and (c) keeps the rest (= non-transferable credit = initial credit – transferable credit) to herself. Thus, by ignoring non-transferable credits, we can focus on credit transfer among coauthors according to the coauthor order, which is represented in numbers without parentheses in the diagonal of the credit mobility matrix.

In addition, the model entails that any first author, as she has no preceding author in coauthor order, gets the transferable credit allocated to herself. This self-allocation of transferable credit, which represents the first author role, is depicted by the self-loop in the coauthorship network (see the black-colored, self-looped tie in Figure 4) in the same way as the non-transferable credit allocated to authors is depicted (see the gray-colored, self-looped ties in Figure 4). In Table 10, the credits self-transferred to first authors are shown in parentheses. Thus, the cell value without parentheses minus that with parentheses represents the credits transferred among authors. Table 11 normalized the values shown in Table 10 by dividing a value(s) in each cell by the Block size: i.e., Block1 size = 175, Block2 size = 257 and Block3 size = 429.

Table 10: Coauthorship Credit Transfer with Non-Transferable Credit Excluded

|        | Block1           | Block2           | Block3         |
|--------|------------------|------------------|----------------|
| Block1 | 76.45 (45.40)    | 8.72             | 2.43           |
| Block2 | 8.62             | 25.13 (22.29)    | 0.90           |
| Block3 | 18.38            | 16.35            | 9.43 (1.99)    |

Table 11: Coauthorship Credit Transfer with Non-Transferable Credit Excluded (Normalized)

|        | Block1           | Block2           | Block3           |
|--------|------------------|------------------|------------------|
| Block1 | 0.4369 (0.2594)  | 0.0498           | 0.0139           |
| Block2 | 0.0335           | 0.0978 (0.0867)  | 0.0035           |
| Block3 | 0.0428           | 0.0381           | 0.0220 (0.0046)  |

The grey-shaded diagonals in Table 10 and 11 represent the within-group transfer and the self-allocation (values in parentheses) of coauthorship credit. Contrary to our expectations, the indegree flow does not follow the '20:80' distribution. In other words, the top 20% of scholars did not obtain 80% of the credits. Instead, the differences between the sizes of self-allocated credit for first authors in each block are noticeable: an average of 0.2594 credits in Block1 was assigned by authors to themselves due to their role as the first author. This value is almost three times larger than the 0.0867 in Block2 and 56 times larger than the 0.0046 in Block3. Thus, we can say that, on average, authors in Block1 led collaboration as the first author much more often and/or with fewer collaborators than others in Block2 and 3. This also means that the first author role is not evenly assigned: a relatively small number of authors repeatedly take the lead in (small-sized) collaboration.

The flow of credits between Blocks should also be highlighted. Scholars in Block3, who account for half (= 429) of 861 unique authors in this dataset, transferred an average 0.0428 credits to Block1 and 0.0318 to Block2. In contrast, the incoming transfer to Block 3 was 0.0139 from Block1 and 0.0035 from Block2. This asymmetric exchange of credits means that half of scholars (= Block3 = the lower 50%) participated in collaboration usually as secondary authors to those in the upper 50% (= Block1 and Block2).

Interestingly, Block1 authors transferred slightly more credits to Block2 authors (= 0.0498) than vice versa (Block2 gave 0.0335 to Block1). Block1 authors also supported Block3 authors as secondary authors, although the size of credits transferred by Block1 authors to Block3 authors (=0.0139) is only 32% of the 0.0428 credits transferred by Block3 authors to Block1 authors. Meanwhile, Block2 authors received 0.0381from Block3 authors and returned only 9% (= 0.0035). Thus, we can say that the top 175 authors in Psychometrika not only led coauthoring more often and/or with fewer collaborators than others in Block2 and Block3, but also support collaboration as secondary authors than Block2 authors do[5].

Another noticeable feature is that Block1 authors transfer more credits to their Block members than to authors in other Blocks. For example, authors in Block1 transferred an average 0.1775 credits (= 0.4369 – 0.2594) to within-block members, while they sent 0.0498 to Block2 members and 0.0139 to Block3 members. In other words, 74% of transferable credits (except the self-allocated credits for first authors) of Block1 members were exchanged among within-block members. In contrast, authors in Block 2 and Block3 transferred more to members outside their blocks than to within-block members. For example, Block2 authors transferred only 0.0111 (=0.0978 – 0.0867) among within-block authors, while they sent 0.0370 (= 0.0335 + 0.0035) to authors in Block1 and 2. Block3 authors exchanged 0.0174 (= 0.0220 – 0.0046), while they sent 0.0809 (= 0.0428 + 0.0381) to out-block members. This implies that prestigious authors in terms of indegree supported prestigious colleagues by being their secondary authors.

By observing these credit transfers within and between blocks of authors, we can infer the hierarchy of collaboration in Psychometrika. Authors in each block show distinct patterns of collaboration. This illustrates that the newly introduced directed coauthorship network approach enables a more detailed interpretation of collaboration among scholars, which could be missed when solely relying on undirected networks.

*Structural vs. Real-world Prestige Validation*

Clearly, prestige measure in this study should not be regarded as a true reflection of real-world prestige. The conceptualization of prestige in this paper is defined in terms of ties received and is therefore purely based on network structure. This notion of structural prestige, like other centrality measures, can take on a different meaning from the notion that people have when they think of prestigious scholars in the actual world (de Nooy et al., 2011; Knoke & Burt, 1983). To check the validity of new bibliometrics measures, several studies validate their measures by comparing the results per measure against a pool of scholars assumed to be prominent: e.g., membership in conference committees, editorial boards, and serving as keynote speakers in conferences (Liu et al., 2005).

We follow this strategy by comparing the results per measure against a pool of scholars assumed to be prominent in the Psychometric Society and its official journal Psychometrika. For this purpose, we selected scholars who have served as presidents of the society between 1991 and 2014 (elected), program committee members for the 2009-2013 period, trustees of the society from 2010, editorial council members from 2010, associate editors in 2013, and Lifetime Achievement Winners awarded by the society from 2008 to 2012. A total of 54 scholars' names were collected. From this set, we eliminated those who have not published in Psychometrika, which reduced the number of individuals to 41. We matched these 41 authors individually against the top 100 authors based on rankings results from centrality and prestige measures.

The results are shown in Figure 6. Among the top 100 authors ranked by the number of collaborators (degree), 19 of them also occur in the set of real world prominent scholars. The betweenness centrality outperformed degree centrality by matching 25 prominent scholars. With indegree prestige ranking, 27 authors from that pool were represented. These outcomes indicate that, in the Psychometric Society and the Psychometrika journal, the indegree prestige measure can detect real-world prominent scholars as good as betweenness centrality or better than degree and closeness centralities[6].

Figure 6: Detection of Real-World Prominent Scholars Based on Prominence Measures

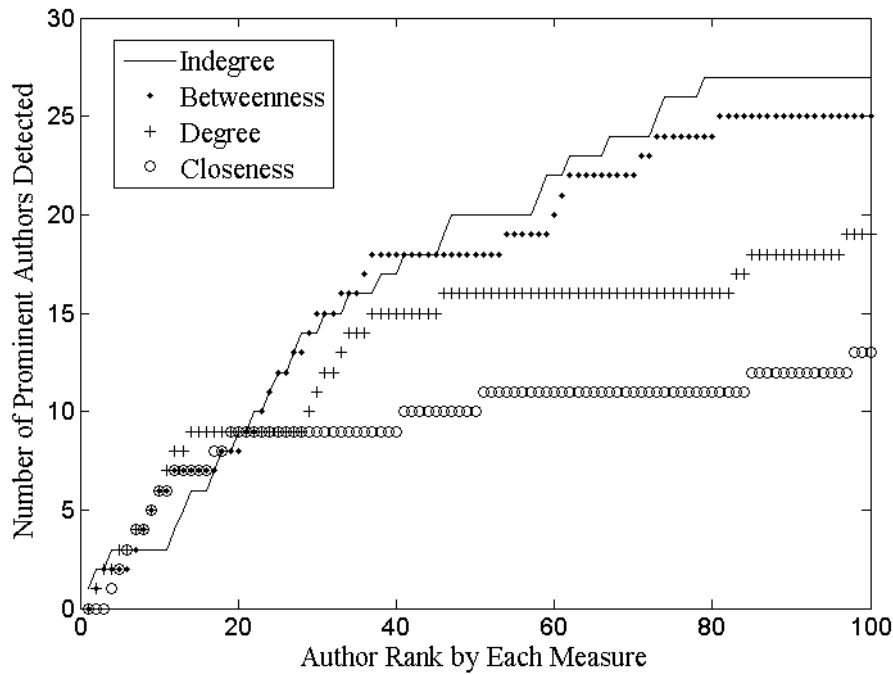

**Conclusion and Discussion**

In this paper, we applied a directed, weighted and self-looped network model of coauthorship credit allocation proposed by Kim and Diesner (2014) to an analysis of a coauthorship network. This method enabled us to consider the coauthor order, which has not been fully conceptualized in previous coauthorship network research. Through the proposed model, coauthors are connected via directed relations of transferring credits assigned to them according to the coauthor order and the number of collaborators per paper. This adopted approach allows us to use the indegree prestige measure, which was applied to a coauthorship network of the Psychometrika scholars for illustration of the measure's usage. We showed that this measure can contribute to a deeper understanding of coauthoring patterns such as the hierarchical structure of collaboration. The validity of the proposed measure was also tested against a real world group of prominent scholars in the considered field and compared with other centrality measures.

This study has limitations that apply to the data and methodology. We used real-world data from Psychometrika. This decision was based on our belief that this particular journal could serve as a good example, mainly because it is an outlet for high quality research in the field of psychology where the coauthor ordering convention and perceived authorship credit allocation have been well studied. The findings from this study, however, should not be generalized to other journals in psychology or to other fields without empirical validation.

Another limitation is that most scholars publish in more than one journal. Thus, the analysis of a single journal would show an incomplete map of collaborative activities among scholars. This might explain why some prominent scholars in terms of president leadership, committee and editor membership did not appear in Psychometrika or were not ranked highly based on the four prominence metrics. Analogously, additional prominent scholars might have not been detected because they usually publish as a single author, which were excluded in this study.

In addition, it is undeniable that the choice of a pool of scholars who were matched against for validating measurement outcomes might be biased. Since there is no such thing as a widely-agreed-upon definition or ranking of scholar prominence, a biased decision is inevitable. For example, if the program committee members of the Psychometric Society before 2009 were included, the result of matching real world prominent scholars and structurally prominent ones might be different.

Most importantly, several assumptions of the new model may well be challenged. The structural prestige constructed through a directed coauthorship network is based on the perception of readers of research products, not on the actual contribution of coauthors. Moreover, the coauthorship credit shares assigned to an author based on the order and number of coauthors can only approximate the empirical data. Although the model can fit coauthorship credit distribution to real-world's perceived allocation with a distribution factor, it is not flexible enough to represent every possible authorship credit distribution. It is inevitable that a difference between the empirical data and the fitted model scores can exist.

Despite the outlined limitations, this study is meaningful in that the proposed conceptualization is a step towards revealing diverse aspects of coauthor relationship. For example, by relaxing the assumption that each paper is assigned an equal value, we can assign the number of citation to each paper and, thus, integrate co-author order and citation in coauthorship network analysis. In addition, single-authored papers, which have been excluded in traditional coauthorship studies, can be considered for scholarly impact as the model can conceptualize them as self-looped networks[7]. Furthermore, based on the directed network approach, coauthorship can be studied with other previously well-defined network analysis techniques such as position and role, hierarchical clustering and network topology. We hope this study will invite other researchers to develop more refined and sophisticated methods and investigate diverse research questions on scholarly collaboration using the directed coauthorship network concept.

**Acknowledgments**

The authors would like to thank Li Tong for providing a help to parse the data and Andrew Higgins for editing the manuscript. We also would like to thank anonymous reviewers who helped us to improve our paper with their insightful comments.

**Footnotes**

1. Although the last-positioned corresponding author is likely to contribute most to a paper in most cases, we cannot be absolutely sure for psychology due to the lack of research directly addressing the issue. Thus, we conducted a sensitivity test for three scenarios: the last –positioned corresponding author contributes (1) more than (2) less than and (3) equal to the first author. The indegree prestige rankings (Kendall's tau rank order correlation) from these scenarios were highly correlated with one another ($\tau = 0.97$~$0.99$, $p < 0.01$).

2. Ideally, the ground-truth data would be a dataset of contribution allocation directly assigned by authors of target papers. But it would be very difficult to invite all authors to report their contribution. Even if that is possible, authors in a paper might have different opinions on their contribution to a work (Shapiro, Wenger, & Shapiro, 1994). We use the perception of peer scholars on author contribution, following previous studies on coauthorship credit allocation (Hagen, 2010; Maciejovsky, Budescu, & Ariely, 2008; Wren et al., 2007). The overarching assumption of those studies is that coauthorship credit allocation is more of an issue to evaluators than to authors themselves: for example, peer scholars interested in the same field, journal editors, or committee members in charge of assessing scholars for inviting, hiring, promotion, tenure, or funding. Another assumption seems to be that the respondents of perceived coauthorship credit allocation studies were likely to be authors of their own papers in the past and, thus, their responses could be regarded as reflecting their evaluation of coauthoring experiences. Thus, we assume that the perceived coauthorship credit allocation can be used for estimating the contribution of each author. But, considering its aforementioned limitation, we regard it as a proxy, an approximate reflection of coauthor contribution in real world.

3. The coauthor lists here were rearranged to reflect the corresponding author position.

4. When a paper is assumed to have one value, the sum of all credits is 663, being equal to the total number of multi-authored papers in Psychometrika.

5. This can be explained in another way. Authors might increase their indegree prestige by repeatedly collaborating with a small number of coauthors as lead authors or exchange the leading author role among them. Whether prestigious scholars collaborate with prestigious others or authors' strategic collaboration makes them prestigious is another sociological issue beyond the scope of this study.

6. A paired-samples t-test was conducted to evaluate the difference between indegree prestige ($M = 17.91$, $SD = 8.31$) and betweenness centrality ($M = 17.09$, $SD = 7.38$) in detecting prominent scholars. The difference was statistically significant, $t(99) = 6.04$, $p < .0005$ (two-tailed). The mean difference was .82 with a 95% confidence interval ranging from .55 to 1.09. The eta squared statistic, 0.27, indicated a large effect size. Therefore, we can say that indegree prestige performs better than degree centrality. This test was the idea of an anonymous reviewer.

7. The idea of assigning a citation frequency as a paper value and extending the model to single-authored papers was suggested by another reviewer.

**References**

Barabasi, A. L., Jeong, H., Neda, Z., Ravasz, E., Schubert, A., & Vicsek, T. (2002). Evolution of the social network of scientific collaborations. *Physica A-Statistical Mechanics and Its Applications, 311*(3-4), 590-614. doi: 10.1016/s0378-4371(02)00736-7

Beasley, B. W., & Wright, S. M. (2003). Looking forward to promotion: Characteristics of participants in the prospective study of promotion in academia. *Journal of General Internal Medicine*, *18*(9), 705-710. doi: 10.1046/j.1525-1497.2003.20639.x

Borgatti, S. P. (2006). Identifying sets of key players in a social network. *Computational & Mathematical Organization Theory, 12(1),* 21-34. doi: 10.1007/s10588-006-7084-x

Chan, K. C., Chen, C. R., & Cheng, L. T. W. (2009). Co-authorship patterns in accounting research. *Internation Review of Accounting, Banking and Finance, 1(2),* 1-10.

Clauset, A., Shalizi, C. R., & Newman, M. E. (2009). Power-law distributions in empirical data. *SIAM review, 51(4)*, 661-703.

De Stefano, D., Giordano, G., & Vitale, M. P. (2011). Issues in the analysis of co-authorship networks. *Quality & Quantity, 45(5)*, 1091-1107. doi: 10.1007/s11135-011-9493-2

Ding, Y. (2011). Applying weighted PageRank to author citation networks. *Journal of the American Society for Information Science and Technology, 62(2),* 236-245. doi: 10.1002/Asi.21452

Endersby, J. W. (1996). Collaborative research in the social sciences: Multiple authorship and publication credit. *Social Science Quarterly, 77(2),* 375-392.

Faust, K. (1997). Centrality in affiliation networks. *Social Networks, 19(2),* 157-191. doi: 10.1016/s0378-8733(96)00300-0

Fiala, D., Rousselot, F., & Jezek, K. (2008). PageRank for bibliographic networks. *Scientometrics, 76(1),* 135-158. doi: 10.1007/s11192-007-1908-4

Field, A. (2009). *Discovering statistics using SPSS*. Thousand Oaks, CA: Sage Publications Limited

Freeman, L. C. (1978). Centrality in social networks: Conceptual clarification. *Social Networks, 1(3),* 215-239.

Galam, S. (2011). Tailor based allocations for multiple authorship: a fractional gh-index. *Scientometrics, 89(1),* 365-379. doi: 10.1007/s11192-011-0447-1

Goyal, S., van der Leij, M. J., & Moraga-Gonzalez, J. L. (2006). Economics: An emerging small world. *Journal of Political Economy, 114(2)*, 403-412. doi: 10.1086/500990

Hagen, N. T. (2010). Harmonic publication and citation counting: sharing authorship credit equitably–not equally, geometrically or arithmetically. *Scientometrics, 84(3),* 785-793.

He, B., Ding, Y., & Yan, E. (2012). Mining patterns of author orders in scientific publications. *Journal of Informetrics, 6(3)*, 359-367. doi: 10.1016/j.joi.2012.01.001

Jennings, M. M., & El-adaway, I. H. (2012). Ethical issues in multiple-authored and mentor-supervised publications. *Journal of Professional Issues in Engineering Education and Practice, 138(1),* 37-47. doi: 10.1061/(asce)ei.1943-5541.0000087

Jian, D., & Xiaoli, T. (2013). Perceptions of author order versus contribution among researchers with different professional ranks and the potential of harmonic counts for encouraging ethical co-authorship practices. *Scientometrics, 96(1)*, 277-295.

Kim, J., & Diesner, J. (2014). A network-based approach to coauthorship credit allocation. *Scientometrics*, 1-16. doi: 10.1007/s11192-014-1253-3

Knoke, D., & Burt, R.S. (1983). Prominence. In R.S. Burt & M.J. Miner (Eds.), *Applied network analysis: A methodological introduction* (pp. 195-222). Beverly Hills, CA: SAGE.

Laband, D. N., & Tollison, R. D. (2006). Alphabetized coauthorship. *Applied Economics, 38(14),* 1649-1653. doi: 10.1080/00036840500427007

Liu, X. M., Bollen, J., Nelson, M. L., & Van de Sompel, H. (2005). Co-authorship networks in the digital library research community. *Information Processing & Management, 41(6),* 1462-1480. doi: 10.1016/j.ipm.2005.03.012

Li, Y. J., & You, C. (2013). What is the difference of research collaboration network under different projections: Topological measurement and analysis. *Physica a-Statistical Mechanics and Its Applications, 392(15),* 3248-3259. doi: 10.1016/j.physa.2013.03.021

Maciejovsky, B., Budescu, D. V., & Ariely, D. (2008). The researcher as a consumer of scientific publication: How do name-ordering conventions affect inferences about contribution credits? *Marketing Science, 28(3),* 589-598. doi: 10.1287/mksc.1080.0406

Marušić, A., Bošnjak, L., & Jerončić, A. (2011). A systematic review of research on the meaning, ethics and practices of authorship across scholarly disciplines. *Plos One, 6(9),* e23477.

Mattsson, P., Sundberg, C. J., & Laget, P. (2011). Is correspondence reflected in the author position? A bibliometric study of the relation between corresponding author and byline position. *Scientometrics, 87(1)*, 99-105. doi: 10.1007/s11192-010-0310-9

Milojević, S. (2012). How Are Academic Age, Productivity and Collaboration Related to Citing Behavior of Researchers? *Plos One, 7(11)*. doi: DOI 10.1371/journal.pone.0049176

Moody, J. (2004). The structure of a social science collaboration network: Disciplinary cohesion from 1963 to 1999. *American Sociological Review, 69(2),* 213-238.

Newman, M. E. J. (2001). The structure of scientific collaboration networks. *Proceedings of the National Academy of Sciences of the United States of America, 98(2)*, 404-409. doi: 10.1073/pnas.021544898

Newman, M. E. J. (2004). Coauthorship networks and patterns of scientific collaboration. *Proceedings of the National Academy of Sciences of the United States of America, 101*, 5200-5205. doi: 10.1073/pnas.0307545100

Prell, C. (2012). *Social network analysis: History, theory & methodology*. Thousand Oaks, CA: SAGE.

Retzer, V., & Jurasinski, G. (2009). Towards objectivity in research evaluation using bibliometric indicators - A protocol for incorporating complexity. *Basic and Applied Ecology, 10(5),* 393-400. doi: 10.1016/j.baae.2008.09.001

Riesenberg, D., & Lundberg, G. D. (1990). The order of coauthorship - Who's on 1st. *JAMA-Journal of the American Medical Association, 264(14)*, 1857-1857. doi: 10.1001/jama.264.14.1857

Rodriguez, M. A., & Pepe, A. (2008). On the relationship between the structural and socioacademic communities of a coauthorship network. *Journal of Informetrics, 2(3)*, 195-201. doi: 10.1016/j.joi.2008.04.002

Shapiro, D. W., Wenger, N. S., & Shapiro, M. F. (1994). The contributions of authors to multiauthored biomedical research papers. *JAMA-Journal of the American Medical Association-International Edition, 271(6)*, 438-442.

Shumate, M., Pilny, A., Atouba, Y., Kim, J., Peña-y-Lillo, M., Copper, K. R., Sahagun, A., & Yang, S. (2013). A taxonomy of communication networks. In E. L. Cohen (Ed.), *Communication Yearbook 37* (pp. 94-123). New York, NY: Routledge.

Sidiropoulos, A., & Manolopoulos, Y. (2006). Generalized comparison of graph-based ranking algorithms for publications and authors. *Journal of Systems and Software, 79(12),* 1679-1700. doi: 10.1016/j.jss.2006.01.011

Spiegel, D., & Keith-Spiegel, P. (1970). Assignment of publication credits: Ethics and practices of psychologists. *American Psychologist, 25(8),* 738-&. doi: 10.1037/H0029769

Thomas, P. A., Diener-West, M., Canto, M. I., Martin, D. R., Post, W. S., & Streiff, M. B. (2004). Results of an academic promotion and career path survey of faculty at the Johns Hopkins University School of Medicine. *Academic Medicine, 79(3),* 258-264. doi: 10.1097/00001888-200403000-00013

Tscharntke, T., Hochberg, M. E., Rand, T. A., Resh, V. H., & Krauss, J. (2007). Author sequence and credit for contributions in multiauthored publications. *Plos Biology, 5(1),* 13-14. doi: 10.1371/journal.pbio.0050018

Vinkler, P. (1993). Research contribution, authorship and team cooperativeness. *Scientometrics, 26(1),* 213-230.

Wagner, M. K., Dodds, A., & Bundy, M. B. (1994). Psychology of the scientist : Assignment of authorship credit in psychological-research. *Psychological Reports, 74(1),* 179-187.

Waltman, L. (2012). An empirical analysis of the use of alphabetical authorship in scientific publishing. *Journal of Informetrics, 6(4),* 700-711. doi: 10.1016/j.joi.2012.07.008

Wasserman, S., & Faust, K. (1994). *Social network analysis: Methods and applications*. New York, NY: Cambridge University Press.

West, J. D., Jensen, M. C., Dandrea, R. J., Gordon, G. J., & Bergstrom, C. T. (2013). Author-level Eigenfactor metrics: Evaluating the influence of authors, institutions, and countries within the social science research network community. *Journal of the American Society for Information Science and Technology, 64(4),* 787-801. doi: 10.1002/Asi.22790

Wray, K. B. (2002). The epistemic significance of collaborative research. *Philosophy of Science, 69(1),* 150-168. doi: 10.1086/338946

Wren, J. D., Kozak, K. Z., Johnson, K. R., Deakyne, S. J., Schilling, L. M., & Dellavalle, R. P. (2007). The write position: A survey of perceived contributions to papers based on byline position and number of authors. *Embo Reports, 8(11),* 988-991. doi: 10.1038/sj.embor.7401095

Yan, E., & Ding, Y. (2009). Applying centrality measures to impact analysis: A coauthorship network analysis. *Journal of the American Society for Information Science and Technology, 60(10),* 2107-2118. doi: 10.1002/Asi.21128

Yan, E., & Ding, Y. (2011). Discovering author impact: A PageRank perspective. *Information Processing & Management, 47(1)*, 125-134. doi: 10.1016/j.ipm.2010.05.002

Yin, L. C., Kretschmer, H., Hanneman, R. A., & Liu, Z. Y. (2006). Connection and stratification in research collaboration: An analysis of the COLLNET network. *Information Processing & Management, 42(6),* 1599-1613. doi: 10.1016/j.ipm.2006.03.021